\RequirePackage{fix-cm}
\documentclass[twocolumn]{svjour3}          
\smartqed
\usepackage{graphicx}
  
\usepackage{microtype}

\usepackage{siunitx}
\DeclareSIUnit{\belmilliwatt}{Bm}
\DeclareSIUnit{\dBm}{dBm}
\DeclareSIUnit{\dB}{dB}
\usepackage{xcolor}
\usepackage{authblk}

\usepackage{hyperref}
\hypersetup{colorlinks=true,linkcolor=blue, allcolors=blue}

\usepackage[misc]{ifsym}

\usepackage{multibib}
\newcites{Appendix}{Appendix References}

\usepackage{doi}
\usepackage{amsmath}

\usepackage{cite}

\usepackage[title]{appendix}

\def\makeheadbox{{%
\hbox to0pt{\vbox{\baselineskip=10dd\hrule\hbox
to\hsize{\vrule\kern3pt\vbox{\kern3pt
\hbox{This is a pre-print of an article published in Nature Communications. \doi{10.1038/s41467-020-20591-5}.}
\kern3pt}\hfil\kern3pt\vrule}\hrule}%
\hss}}}

\begin{document}

\title{Point-to-Point Stabilised Optical Frequency Transfer with Active Optics}

\author{Benjamin P. Dix-Matthews$^{1,2}$, Sascha W. Schediwy$^{1,2}$, David R. Gozzard$^{1,2}$, Etienne~Savalle$^{3}$, Fran\c{c}ois-Xavier Esnault$^{4}$, Thomas L\'{e}v\`{e}que$^{4}$, Charles~Gravestock$^{1}$, Darlene~D'Mello$^{1}$, Skevos Karpathakis$^{1}$, Michael Tobar$^{2}$, Peter~Wolf$^{3}$
}

\authorrunning{B. P. Dix-Matthews et al.} 

\institute{
    \begin{enumerate}
    \item [\Letter] Benjamin P. Dix-Matthews \\ benjamin.dix-matthews@research.uwa.edu.au\\
    \item [$^1$~] International Centre for Radio Astronomy Research, The University of Western Australia, Perth, Australia
    \item [$^2$~] Australian Research Council Centre of Excellence for Engineered Quantum Systems, The University of Western Australia, Perth, Australia
    \item [$^3$~] SYRTE, Observatoire de Paris, Universit\'e PSL, CNRS, Sorbonne Universit\'e, LNE, Paris, France
    \item [$^4$~] Centre National d'\'{E}tudes Spatiales (CNES), Toulouse, France
    \end{enumerate}
}

\date{Compiled on: \today}

\maketitle

\begin{abstract}
Timescale comparison between optical atomic clocks over ground-to-space and terrestrial free-space laser links will have enormous benefits for fundamental and applied science, from measurements of fundamental constants and searches for dark matter, to geophysics and environmental monitoring.
However, turbulence in the atmosphere creates phase noise on the laser signal, greatly degrading the precision of the measurements, and also induces scintillation and beam wander which cause periodic deep fades and loss of signal.
We demonstrate phase stabilized optical frequency transfer over a \SI{265}{\meter} horizontal point-to-point free-space link between optical terminals with active tip-tilt mirrors to suppress beam wander, in a compact, human-portable set-up. A phase stabilized \SI{715}{\meter} underground optical fiber link between the two terminals is used to measure the performance of the free-space link. The active optics terminals enabled continuous, coherent transmission over periods of up to an hour. We achieve an \SI{80}{\dB} suppression of atmospheric phase noise to \SI{3d-6}{\square\radian\per\hertz} at \SI{1}{Hz}, and an ultimate fractional frequency stability of $1.6\times10^{-19}$ after~\SI{40}{\second} of integration. At high frequency this performance is limited by the residual atmospheric noise after compensation and the frequency noise of the laser seen through the unequal delays of the free space and fiber links. Our long term stability is limited by the thermal shielding of the phase stabilization system. We achieve residual instabilities below those of the best optical atomic clocks, ensuring clock-limited frequency comparison over turbulent free-space links.


\end{abstract}

\section{Introduction}

Modern optical atomic clocks have the potential to revolutionise high-precision measurements in fundamental and applied sciences \cite{grotti2018geodesy,riehle2017optical,delva2017test,lisdat2016clock,takano2016geopotential,yamaguchi2011direct,rosenband2008frequency}. The ability to realise remote timescale comparison in situations where fiber links are impractical or impossible, specifically, between ground- and space-based optical atomic clocks \cite{kang2019free,swann2019measurement,bergeron2019femtosecond,sinclair2019femtosecond,Gozzard2018,Sinclair2018,chen2017sub,belmonte2017effect,deschenes2016synchronization,sinclair2016synchronization,bergeron2016tight,Robert2016,laas2015accuracy,Giorgetta2013,Djerroud:s}, will enable significant advances in fundamental physics and practical applications including tests of the variability of fundamental constants \cite{godun2014frequency,dzuba2000atomic}, general relativity \cite{Delva2018,altschul2015quantum}, searches for dark matter \cite{derevianko2014hunting}, geodesy \cite{mcgrew2018atomic,mehlstaubler2018atomic,Lion2017,Denker2017,flury2016relativistic,huntemann2014improved,bondarescu2012geophysical}, and global navigation satellite systems \cite{lewandowski2011gnss} amongst others \cite{muller2018high,delva2017clocks,clivati2017vlbi,dzuba2016strongly,arzoumanian2015nanograv,takamoto2015frequency,guerlin2015some,uzan2011varying,chou2010optical,wolf2009quantum,lorimer2008binary}. These efforts lead on from optical timing links developed for timescale comparison between microwave atomic clocks \cite{samain2011time,schreiber2010ground,schreiber2009european}, and efforts are underway to develop optical clocks that can be deployed on the International Space Station \cite{bongs2015development} and on dedicated spacecraft \cite{liu2018orbit}.

Similarly, timescale comparisons between mobile terrestrial optical clocks \cite{Takamoto2020,grotti2018geodesy,hannig2019towards,koller2017transportable,Hafner2020}, where one or more mobile clocks are able to be deployed and moved over an area of interest, enable ground tests of general relativity and local geopotential measurements for research in geophysics, environmental monitoring, surveying and resource exploration.

Comparison of both ground- and space-based clocks, and mobile terrestrial clocks, requires frequency transfer over free-space optical links. Just as with timescale comparison over optical fiber links, free-space frequency transfer should have residual instabilities better than those of the optical clocks. However, atmospheric turbulence induces much greater phase noise than a comparable length of fiber \cite{Gozzard2018,Robert2016,djerroud2010coherent,Sinclair2014}. In addition, free-space links through the turbulent atmosphere must also overcome periodic deep fades of the signal amplitude due to beam wander and scintillation. When the size of the optical beam is smaller than the Fried scale of the atmospheric turbulence, the centroid of the beam can wander off the detector while, in the case where the beam is larger than the Fried scale, destructive interference within the beam (scintillation) can also result in loss of signal and so loss of timescale synchronisation \cite{swann2017low,Robert2016,fante1975electromagnetic}. These deep fades can occur 10s to 100s of times per second for vertical links between the ground and space, and also on horizontal links on the order of \SI{10}{\kilo\meter}~\cite{sinclair2016synchronization,Gozzard2018}.

One method to overcome deep fades of the signal is to transmit a series of optical pulses from an optical frequency comb and compare them with another optical frequency comb at the remote site \cite{Giorgetta2013}. While deep fades will result in the loss of some pulses, the time and phase information can be reconstructed from the remaining pulses.

Another method to overcome deep fades is to stabilize the spatial noise caused by atmospheric turbulence by active correction of the emitted and received wave front. 
In general, tip-tilt correction is sufficient when using apertures that are small compared to the Fried scale as beam wander will dominate the deep fades. 
For large apertures the effects of scintillation increase and higher-order corrections using adaptive optics may be necessary.
Tip-tilt stabilization of beam wander for comparison of atomic clocks has previously been demonstrated over \SI{12}{\kilo\meter} with \SI{50}{\milli\meter} scale optics \cite{sinclair2016synchronization} and \SI{18}{\kilo\meter} with larger \SI{250}{\milli\meter} telescopes \cite{kang2019free}.

A further practical concern for the deployment of free-space links is the ability of the system to acquire and track a moving object \cite{bergeron2019femtosecond,cossel2017open}. In that case tip-tilt capability is mandatory, and additionally such a system must be robust while also having as low size, weight, and power as possible for ease of deployment in spacecraft, airborne relay terminals, or mobile ground segments.

In this paper we describe phase-stabilized optical frequency transfer via a \SI{265}{\meter} point-to-point free-space link between two portable optical terminals. Both terminals have \SI{50}{\milli\meter} apertures and utilise tip-tilt active optics to enable link acquisition and continuous atmospheric spatial noise suppression. The terminals are human-portable and ruggedized for daily field deployment to demonstrate the suitability for remote optical timescale comparison. The performance of the phase stabilization system was determined using a separate phase-stabilized \SI{715}{\meter} optical fiber link between the two terminals.

The phase-stabilized free-space optical transfer exhibits an \SI{80}{\dB} improvement in phase noise at \SI{1}{\hertz}, down to \SI{1.0d-5}{\square\radian\per\hertz}, compared to the unstabilized optical transmission.
The active spatial stabilization used at each terminal is effective at suppressing beam wander caused by the atmospheric turbulence, allowing continuous, deep-fade free, coherent transmission on the order of~\SI{3d3}{\second}.
The resulting fractional-frequency stability of the phase-stabilized optical transfer reaches $1.6\times10^{-19}$ after~\SI{40}{\second}.
At timescales beyond~\SI{100}{\second} the fractional-frequency stability flattens, which is likely caused by unstabilized temperature fluctuations.

\section{Results}\label{sec:results}
\subsection{Coherent Optical stabilization System}
\begin{figure*}
    \centering
    \includegraphics[width=\linewidth]{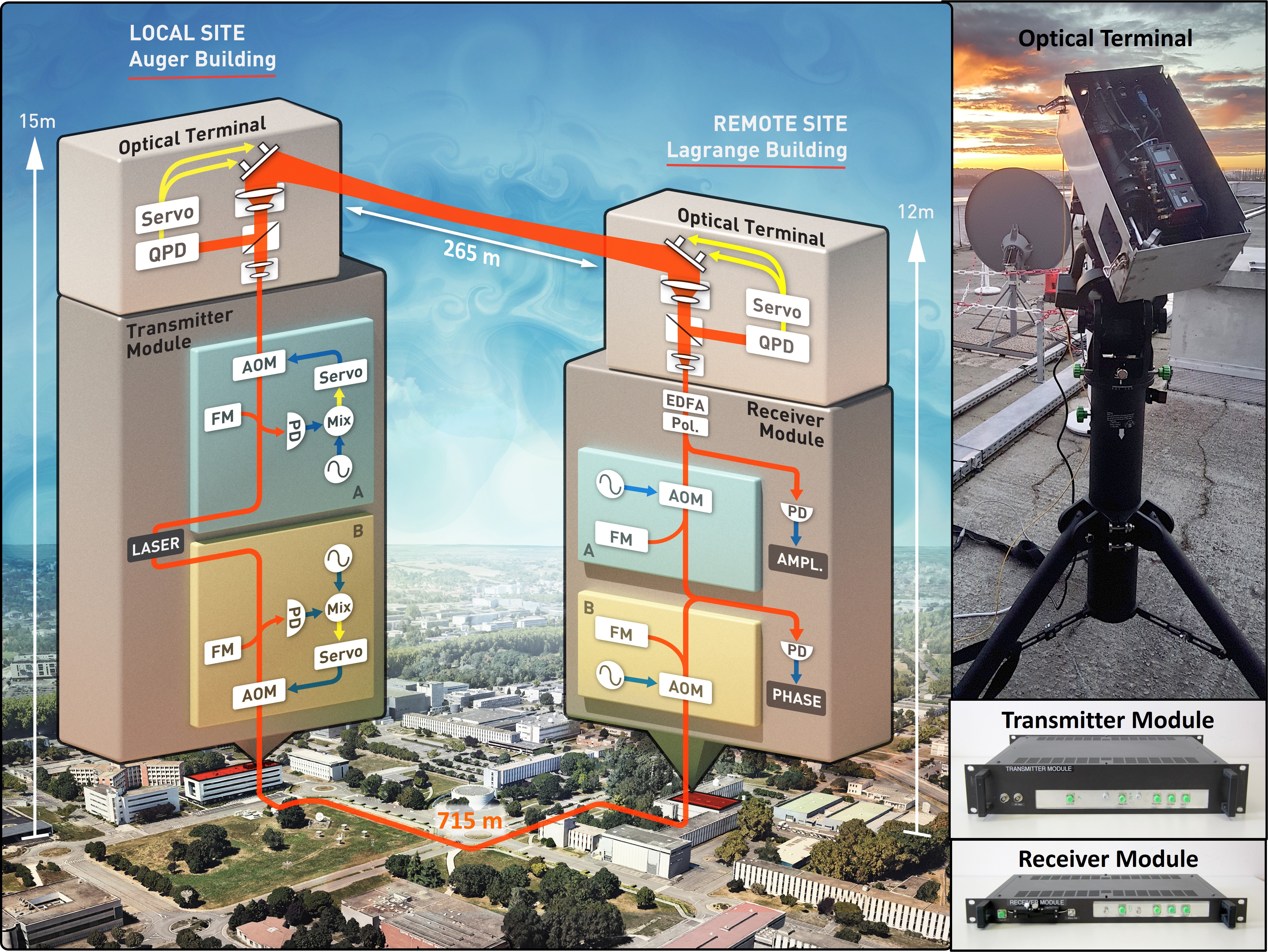}
    \caption{Point-to-point phase stabilized optical frequency transfer between buildings. Two identical phase stabilization systems are implemented across the CNES campus. Both systems have their transmitter located in the Auger building (local site), and both receivers are located in the Lagrange building (remote site). One system transmits the optical signal over a \SI{265}{\meter} free space path between the buildings using tip-tilt active optics terminals while the other transmits via \SI{715}{\meter} of optical fiber. The relative stability of the two optical signals is then measured at the remote site. QPD, quad-photodetector; Pol, polarization controller; PD, photodetector; PLL, phase-locked loop; AOM, acousto-optic modulator; FM, Farady mirror; EDFA, erbium-doped fiber amplifier; Mix, radio frequency electronic mixer.}
    \label{fig:experimentalSystem}
\end{figure*}

Figure~\ref{fig:experimentalSystem} shows the architecture of the phase stabilization systems, as well as the free-space and fiber links used to compare the phase noise performance.

A \SI{15}{\dBm} optical signal from a \SI{1550}{\nano\meter} NKT Photonics X15 Laser was split and passed into two independent phase-stabilization systems, detailed in Section~\ref{sec:phaseSystem}.
One of these phase-stabilized systems operated over the free-space link, and was used to suppress the phase noise resulting from atmospheric turbulence.
The second phase-stabilization system operated over an optical fiber that ran underground between the local and remote sites, and was used to measure the performance of the free-space transmission.

Each side of the free-space link also incorporated tip-tilt active optical terminals (detailed in Section.~\ref{sec:ActiveOpticalTerminals}) that were used to suppress the received optical intensity fluctuations and deep fades caused by beam wander due to atmospheric turbulence.
The remote terminal additionally had a bi-directional optical amplifier that amplified the incoming optical signal (typically by~\SI{\sim13}{\dB}) before passing it to the phase stabilization system, and amplified the reflected portion of the signal for transmission back over the link.

The free-space link spanned \SI{265}{\meter} between two buildings at the Centre National d'\'{E}tudes Spatiales (CNES) campus in Toulouse, as shown in Figure~\ref{fig:experimentalSystem}.
The link passed over grass, sparse trees and roads, and 
was operated during late winter over the course of two weeks. The most favourable conditions were when the sky was overcast and wind speed was low.

\begin{figure*}
    \centering
    \includegraphics[width=\linewidth]{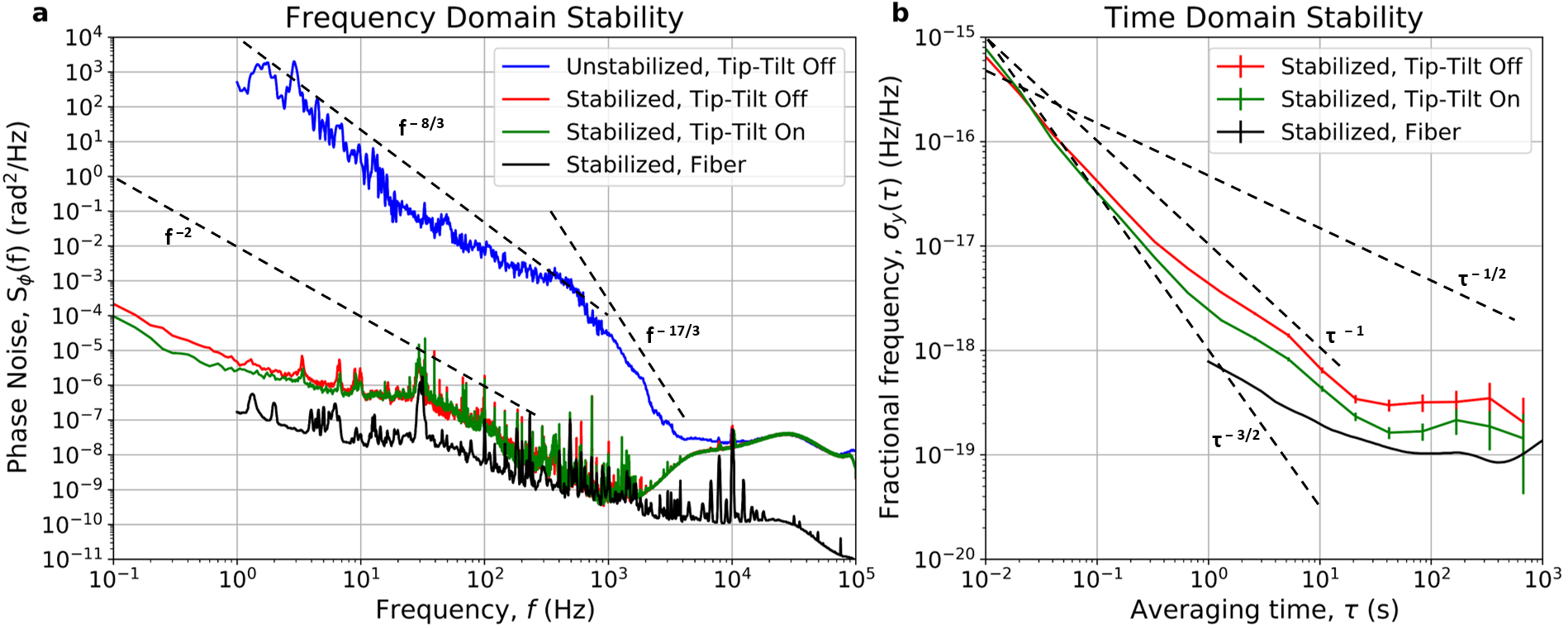}
    \caption{Phase noise PSD, \textbf{\textsf{a}}, and frequency stability (modified Allan deviation), \textbf{\textsf{b}}, of the optical transmission measured at the remote site. Blue trace, free-space link phase stabilization off, tip-tilt active optics off; red trace, free-space link phase stabilization on, tip-tilt active optics off; green trace, free-space link phase stabilization on, tip-tilt active optics on; black trace, system noise floor with both phase stabilization systems transmitting over parallel optical fiber. Dashed lines show key gradients of interest.}
    \label{fig:stabilityResults}
\end{figure*}

\subsection{Fully coherent transfer over a true point to point link}
Figure~\ref{fig:stabilityResults} shows the measurements for the fiber noise floor (black) and phase stabilization off (blue) cases made with a Microsemi 3120A Phase Noise Test Probe.
Phase noise measurements for the phase-stabilized cases with (green) and without (red) tip-tilt were obtained using an Ettus X300 Software Defined Radio operating as a continuous IQ demodulator, and are also shown in Figure~\ref{fig:stabilityResults}.

Further discussion of the measurement equipment architecture and choice may be found in the Supplementary Material~\ref{sec:measVerification}.
The phase noise Power Spectral Densities (PSD) found using the Ettus X300 shows good agreement with the Microsemi 3120A within overlapping frequency ranges, as shown in the Supplementary Material~\ref{sec:measVerification}.

When the phase stabilization and tip-tilt systems are off, the measured noise is expected to be dominated by atmospheric turbulence. In theory~\cite{Conan1995} the corresponding PSD is expected to decrease as $f^{-8/3}$ for low frequencies, before dropping sharply as $f^{-17/3}$ due to the averaging effect of the optical aperture. The slopes of our measured PSD are compatible with that model. The transition frequency between the two regimes is given in \cite{Conan1995} by $f_c = 0.3\,V/D$, where $V$ is the transverse wind speed and $D$ the aperture diameter. This is not confirmed in our data, as wind speeds were no more than a few tens of m/s and our beam diameter was about \SI{34}{\milli\meter}. The corresponding theoretical transition frequency is significantly lower than the $\approx$ \SI{400}{\hertz} visible in Figure~\ref{fig:stabilityResults}. We attribute that discrepancy mainly to the fact that the theoretical calculations in \cite{Conan1995} were done for a plane wave impinging on a circular aperture, while our beam is Gaussian and smaller than the receiving aperture. We leave a detailed discussion of that point for future studies, but note that discrepancies between the theoretical model and experimental measurements have been reported previously (see e.g. Tab.~I in \cite{Sinclair2014}).   

When the stabilization system is turned on, we see around eight orders of magnitude reduction in phase noise PSD at \SI{1}{\hertz}, down to \SI{3d-6}{\square\radian\per\hertz}. Having the active tip-tilt terminal engaged appears to offer a slight improvement in phase-stability. At frequencies above roughly \SI{2}{\kilo\hertz} the phase noise performance is limited by the residual phase noise of the laser (this is discussed in Supplementary Material~\ref{sec:noiseFloor}), which also affects the unstabilized measurement above $\approx$\SI{10}{\kilo\hertz}. At lower frequency (roughly \SIrange{200}{2000}{\hertz}) we are most likely limited by the noise floor resulting from the operation of our compensation system when applied to the atmospheric phase noise, as shown in detail in Supplementary Material~\ref{sec:noiseFloor}.

\begin{figure*}
    \centering
    \includegraphics[width=\linewidth]{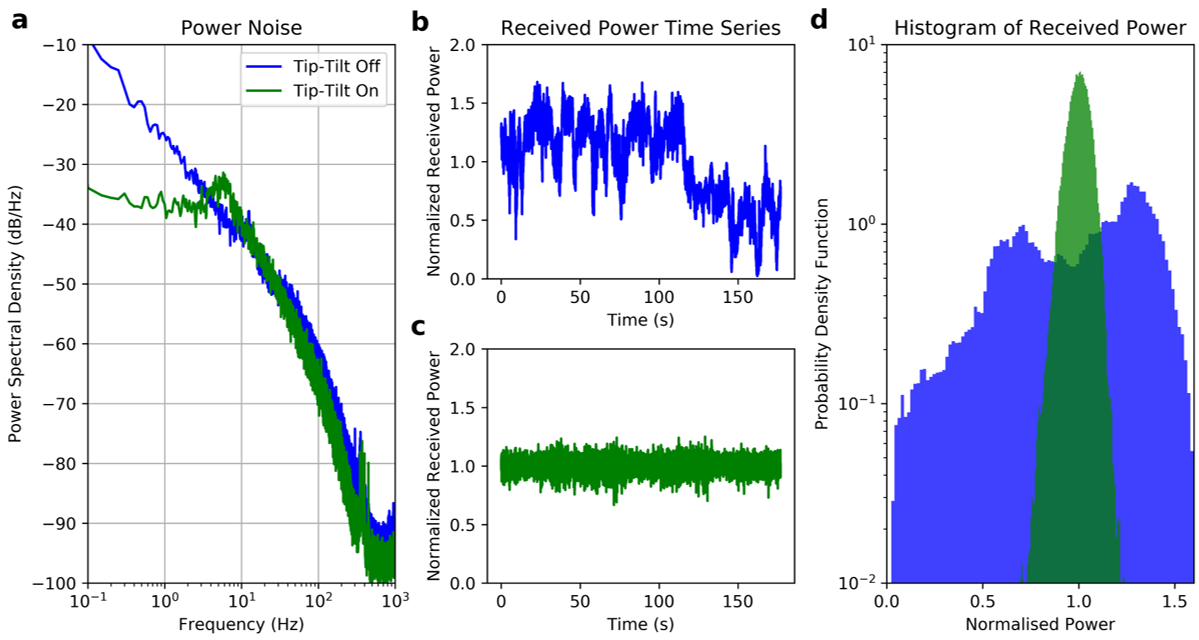}
    \caption{Normalized Power $(P/\langle P\rangle)$ of the free-space optical signal received at the remote site with tip-tilt active optics off (blue) and on (green) over three minutes. \textbf{\textsf{a}} shows the power spectral density, \textbf{\textsf{b}} and \textbf{\textsf{c}} show the normalized power time series for the tip-tilt off and tip-tilt on cases respectively, and \textbf{d} shows the histogram of the normalized power values.}
    \label{fig:amplitudeStability}
\end{figure*}

The long term fractional frequency stability of the stabilized signals is shown in Figure~\ref{fig:stabilityResults}~(b) in terms of modified Allan deviation (MDEV). This provides an alternative tool for assessing the performance of the stabilized optical transfer, with a particular focus on stability at longer time scales. It was calculated using the same Ettus X300 phase data, after removal of a quadratic fit in phase. The removed linear and quadratic coefficients were $0.15$~rad/s and $6.1\times 10^{-7}$~rad/s$^2$ for the tip-tilt ``on" data ($0.14$~rad/s and $-1.3\times 10^{-7}$~rad/s$^2$ for tip-tilt ``off"). We attribute that variation to the differential optical length change in the uncompensated short ($\sim 60$~cm) fibres between the laser and first beam splitters on the transmitter side, and last beam splitters and photodiode on the receiver side. For example, with a typical temperature coefficient of optical length change $\delta L/L \approx 10^{-5}$/K, the linear term corresponds to a differential temperature variation of order mK/s, very likely on our rooftop location without air conditioning.

The MDEV averages as a combination of $\tau^{-3/2}$ and $\tau^{-1}$ power laws until an integration time of around~\SI{20}{\second}, indicating that the dominant frequency noise at short timescales is white phase and flicker phase noise, in agreement with the phase noise PSD. The optimum stability reached when the active optical terminal was turned off is $3.0\times10^{-19}$ at~\SI{40}{\second} of integration time.

When the active tip-tilt terminal is engaged a slight improvement in stability is seen for integration times longer than~\SI{0.02}{\second} (consistent with the phase noise PSD), and the transfer is made more robust.
This results in an optimum stability of $1.6\times10^{-19}$ reached after~\SI{40}{\second} of integration, a factor of two improvement over the case without active tip-tilt control.


At longer timescales the stability does not integrate down further. This is likely due to long term residual temperature fluctuations in the local and remote sites affecting the uncompensated parts of the two links, as discussed above, and observed in \cite{Gozzard2018}. With better thermal regulation the fractional frequency stability is expected to continue averaging down to a lower limit.

The power of the optical signal received by the remote site was recorded in order to measure the atmospheric induced fluctuations encountered during a one-way pass of the free-space link.
Immediately after the active terminal, a fiber splitter was used to send a small portion of the received signal to a fiber-photo-detector with a linear response to the received optical power.
The response of this detector was then digitized at \SI{4}{\kilo\hertz}. 

Figure~\ref{fig:amplitudeStability} shows the frequency domain power of the received power fluctuations.
Without the active tip-tilt terminal engaged, the power fluctuations drop as roughly $f^{-2}$ at low frequency and $f^{-3/2}$ beyond a few Hz. The tip-tilt active optical terminal improves the stability at frequencies below \SI{4}{\hertz}, with over two orders of magnitude reduction in power fluctuations at \SI{0.1}{\hertz}.
The tip-tilt servo bump at approximately \SI{7}{\hertz} is clearly visible. Beyond that bump there is not a significant difference between having the tip-tilt compensation on or off, as expected. It is interesting to note that the \SI{\sim4}{\hertz} crossing point roughly matches the frequency at which the phase noise PSD on Figure~\ref{fig:stabilityResults}) starts improving for tip-tilt on with respect to off, confirming that the phase noise reduction is related to the reduction in power fluctuations. This also implies that better performance of the adaptive optics system (hence lower power fluctuations) is likely to lead to lower phase noise. The tip-tilt system was based on a commercially available unit and the low bandwidth of the system is due to the low gain setting necessary to mitigate some artefacts in the control system firmware (discussed further in Section~\ref{sec:ActiveOpticalTerminals}).

The time domain plots and the histogram provide additional representations of the effect of the active terminal. Without tip-tilt, the optical power fluctuates significantly and at around \SI{100}{\second} there is a step change in the received power. This was likely due to mechanical movement of the optical terminal, such as mechanisms in the telescope mount suddenly slipping. This step in power can also be seen in the bi-modal distribution of the histogram. When the tip-tilt actuation was activated, step behaviour like this was not observed.

Taking the bi-modal feature due to movement of the optical terminal into account, the histograms for both the tip-tilt on and off cases exhibit a log-normal distribution, as is expected of power fluctuations caused by turbulence-induced beam wander. The case with the tip-tilt system on shows a much narrower distribution in received power, indicting more constant optical power levels delivered to the phase stabilization system. This indicates that the tip-tilt active optics are effective at suppressing power fluctuations caused by atmospheric turbulence or movement of the terminals.

For clarity, the optical power time series traces shown in Figure~\ref{fig:amplitudeStability} are normalized to their own average power level. With the tip-tilt system on, the average optical power received at the remote site was 2.4 times higher than the average power level when the tip-tilt system was off. Critically, with the tip-tilt system on, the optical power does not make significant excursions into lower power values, greatly reducing the chance of a cycle slip in the phase stabilization system.

\section{Discussion}
The transfer of stable optical frequency reference signals over free-space is of particular interest to applications involving ad-hoc transmissions between mobile sites.
A specific example of interest is chronometric geodesy \cite{mcgrew2018atomic,mehlstaubler2018atomic,Lion2017,Denker2017,flury2016relativistic,huntemann2014improved,bondarescu2012geophysical}, where frequency comparisons with a mobile optical atomic clock at different positions over the region of interest provide a direct measurement of the gravitational red-shift caused by changes in gravitational field and height. The requirements are that the transfer system provide sufficiently stable optical transmission so that the uncertainty of the frequency comparison is limited by the uncertainty of the optical atomic clocks themselves, be physically robust and portable, and be light and small enough to allow for easy and rapid set up of the terminals in different locations.

The stabilities of the best lab-based optical atomic clocks are approaching $10^{-18}$ for averaging times on the order of \SI{e3}{\second}~\cite{riehle2017optical,mcgrew2018atomic,brewer2019al+,bothwell2019jila,huntemann2016single,nicholson2015systematic}.
Bothwell et al.~\cite{bothwell2019jila} achieve a stability of $4.8\times 10^{-17}/\sqrt{\tau}$ (represented by the $\tau^{-1/2}$ gradient line in Figure~\ref{fig:stabilityResults}), averaging down to a final systematic uncertainty-dominated stability of $2\times 10^{-18}$ within 10 min.
The stability demonstrated using the system described in this paper surpasses this stability by more than an order of magnitude, ensuring that frequency comparison between optical clocks over a turbulent free-space channel such as this will not be limited by the performance of the phase-stabilized link.

Our system is also designed to be physically robust and portable (as shown in Figure~\ref{fig:experimentalSystem}).
The optical terminals are securely built within a steel enclosure that provides protection during transport and while the link is operational.
The optical fiber-based phase-stabilization systems are also built within 2U and 1U 10'' rack-mounted aluminium enclosures, for the transmitter and receiver modules respectively.
The robustness of the terminals was demonstrated by the fact that they were successfully shipped, via conventional couriers, from Perth, Australia to Toulouse, France without damage or misalignment of the optics. One of the terminals was installed in a telescope dome for the duration of the two-week trial period, while the other terminal was set up on an open rooftop and was removed and reset every day.

The effectiveness of the overall system for establishing ad-hoc ground-to-ground optical links for timescale comparison was limited by the performance of the tip-tilt active optics in terms of their ability to effectively suppress atmospheric turbulence. The performance of the tip-tilt system was limited by the QPD (Quadrant Photo Detector) as discussed further in Section~\ref{sec:ActiveOpticalTerminals}, which largely resulted from the limited sensitivity of the QPD. The lower limit of the QPD's operational detected power range is \SI{-10}{\dBm} (\SI{0.1}{\milli\watt}). Large drops in link power would first affect the QPD, causing the tip-tilt system to lose the beam, resulting in a loss of signal and cycle slips in the phase stabilization system. (The phase stabilization system is capable of operating with less than \SI{1}{\nano\watt} of light returning to the transmitter unit \cite{Dix-Matthews2020Coherent}.) Future work will include upgrades to the tip-tilt system to increase the turbulence suppression bandwidth and steering range, and provide more robust beam steering. We also plan to develop an active optical relay that will allow comparing clocks over much longer baselines (up to $\sim$\SI{100}{\kilo\meter}) when installed on an airborne carrier. As the relay will be moving, active optics are compulsory in such a set-up, and the experiment presented here is a first step in that direction.

\section{Method}
\subsection{Phase stabilization System}\label{sec:phaseSystem}
Two phase-stabilization systems, with very similar architectures, are used to stabilize the free-space and fiber paths.
For simplicity, we assume negligible propagation delay and consider only link noise in this section, but these assumptions are revisited in Supplementary Material~\ref{sec:noiseFloor}.
Equivalent variables relating to the free-space and fiber stabilization systems are identified by superscripts of $fs$ and $fb$ respectively.

The stabilization systems are based on the imbalanced Michelson interferometer design developed by Ma et al.~\cite{ma1994delivering,foreman2007remote}, where the long arm of the interferometer is sent over the link and the short arm is reflected by a Faraday mirror to provide an optical frequency reference.
The frequency of the outgoing optical signal is shifted by a transmission acousto-optic modulator (AOM) with a nominal frequency ($\nu_{tr}$) which may be varied ($\Delta\nu_{tr}$).
The shifted optical signal is then sent over the link.

In the free-space system the signal is passed through the active terminal described in Section.~\ref{sec:ActiveOpticalTerminals} and launched over the free-space link.
The signal then reaches the remote site after picking up link phase noise caused mainly by atmospheric turbulence ($\delta \nu^{fs}$).
This optical signal is received by a second active optical terminal and passed through a bi-directional optical amplifier to offset the signal power lost during transmission.

In the fiber system, the signal is passed through an underground fiber running between the two sites.
The transmitted signal picks up link noise due to mechanical and thermal fluctuations along this fiber ($\delta \nu^{fb}$).

At the remote site, each stabilization system passes their received signal through an anti-reflection AOM ($\nu_{ar}$), before outputting half the signal to the end user ($\nu_{out}$).
The output at the remote site of the free-space stabilization system is given by
\begin{equation}\label{eq:outFs}
    \nu_{out}^{fs}=\nu_{L}+\nu_{tr}^{fs}+\Delta\nu_{tr}^{fs}+\nu_{ar}^{fs}+\delta \nu^{fs} \,
\end{equation}
where $\nu_L$ is the laser frequency, while the output of the fiber stabilization system is given by
\begin{equation}\label{eq:outFb}
    \nu_{out}^{fb}=\nu_{L}+\nu_{tr}^{fb}+\Delta\nu_{tr}^{fb}+\nu_{ar}^{fb}+\delta \nu^{fb}\,.
\end{equation}

The two signals are optically beat together at a photodetector and low-pass filtered to produce a down-converted signal,
\begin{equation}\label{eq:measSignal}
    \nu_{meas}=\nu_{out}^{fs}-\nu_{out}^{fb},
\end{equation}

\noindent used to measure the relative stability of the optical signals reaching the remote site through free-space and fiber. The residual phase noise from the free-space transmission dominates the residual phase noise from the fiber transmission over most of the Fourier frequency range.
The AOM frequencies were chosen so that the measured beat signal ($\nu_{meas}$) was at a nominal frequency of \SI{1}{\mega\hertz}.
An external~\SI{10}{\mega\hertz} signal was shared between the two sites via RF over fiber and provided a common reference for the transmitter oscillators and remote site measurement equipment.

The other half of the signals reaching the remote site are reflected by Faraday mirrors back through the anti-reflection AOMs and back over the free-space and fiber links.
For the free-space link, the return signal also passed back through the bi-directional optical amplifier.
At the local site, the signals returning from the fiber and free-space links pass back through their respective transmission AOMs.
Each system then performs a self-heterodyne measurement by beating the returned signal against the short arm of the Michelson interferometer on a photodetector.
The final electrical beat signal,
\begin{equation}\label{eq:vBeat}
    \nu_{beat}=2\nu_{tr}+2\Delta\nu_{tr}+2\nu_{ar}+2\delta \nu,
\end{equation}

\noindent now contains information about the phase noise picked up during the transmission over the link.

This signal is then mixed with a local oscillator of frequency ($2\nu_{tr}+2\nu_{ar}$) and low-pass filtered in order to extract a DC error signal,
\begin{equation}\label{eq:fsServo}
    \nu_{dc}=2\Delta\nu_{tr}+2\delta \nu,
\end{equation}

\noindent for the phase-locked loop (PLL) that stabilizes the transmission frequency.

The PLL then controls the frequency of the transmission AOM in order to drive this error signal to zero, such that $\Delta\nu_{tr}=-\delta \nu$.
This has the effect of suppressing the link phase noise from the free-space (Eq.~\ref{eq:outFs}) and fiber (Eq.~\ref{eq:outFb}) output signals.

\subsection{Active Optical Terminals}\label{sec:ActiveOpticalTerminals}
The active terminals (Figure~\ref{fig:experimentalSystem}) used at each end of the free-space link were reciprocal and identical.
The optical signal is passed through a fiber to free-space collimator with a $1/e^2$ radius of \SI{1.12}{\milli\meter}.
This is then passed through a 50-50 beam splitter (BS).
Half the optical signal is sent to a beam-dump, and the other half is sent to a 15:1 Gaussian beam expander (GBE) with a \SI{48}{\milli\meter} clear aperture.

The signal from the GBE is reflected off a \SI{48}{\milli\meter} flat mirror with active piezo-electric actuators and launched over the free-space link with a $1/e^2$ radius and divergence of approximately \SI{16.8}{\milli\meter} and \SI{29}{\micro\radian} respectively.

The incoming beam is reflected by the active mirror into the GBE.
The BS then sends half the incoming light to the free-space-to-fiber collimator, and the other half to a Quad Photo Detector (QPD).
This QPD is used to detect first-order spatial fluctuations in the incoming beam.
The measured fluctuations are passed through a Proportional Integral (PI) controller and used to drive the piezo-electric actuators on the active mirror in order to suppress these fluctuations and keep the incoming beam centred on the QPD.
The QPD is positioned so that the optical signal coupled by the collimator into the fiber is maximized when the beam is centred on the QPD.

The QPD and active mirror control system is a commercial off-the-shelf system. The achievable turbulence suppression bandwidth of the system during these tests was limited by the low PI controller gain settings which were necessary to reduce the sensitivity of the tip-tilt system to noise in the QPD when the link optical power dropped below the threshold for effective operation of the QPD. When the optical power dropped below this threshold, the tip-tilt system would attempt to steer to the false beam centroid caused by the detector noise, losing the real beam in the process. The low gain settings prevented the tip-tilt system from steering too far off target before sufficient optical power was restored.

\bibliographystyle{spphys}
\small{\bibliography{ref}}

\newpage
\onecolumn
\renewcommand\thefigure{\thesection.\arabic{figure}}    
\renewcommand\theequation{\thesection.\arabic{equation}} 
\begin{appendices}
\section{Phase noise measurement verification}\label{sec:measVerification}
Phase noise measurements in this work were performed by both a Microsemi 3120A Phase Noise Test Probe and an Ettus X300 Software Defined Radio. The Microsemi 3120A is a purpose-built phase/frequency noise measurement device, but is not able to handle large changes in signal amplitude. Performing a direct I/Q demodulation on the Ettus enabled us to continue making accurate and precise phase noise measurements in spite of large variations in signal amplitude and collect raw data time series at 200~kHz sampling for up to an hour. But the noise performance of the Ettus is less well known than the Microsemi, so to ensure compatibility between the measurements from the two devices, concurrent measurements of link phase noise by the two devices, shown in Figure~\ref{fig:ettusVsMicro}, were made and show that the two devices agree on the measured phase noise. Above a few tens of Hz the Ettus and Microsemi are in almost exact agreement. Below these frequencies some discrepancy between the Ettus and Microsemi results can be seen, possibly as a result of the Microsemi's behaviour when trying to measure signals with large amplitude fluctuations. The discrepancy grows to a maxiumum difference of a factor of 2 at 1~Hz.

\setcounter{figure}{0}
\begin{figure}[ht]
    \centering
    \includegraphics[width=0.6\linewidth]{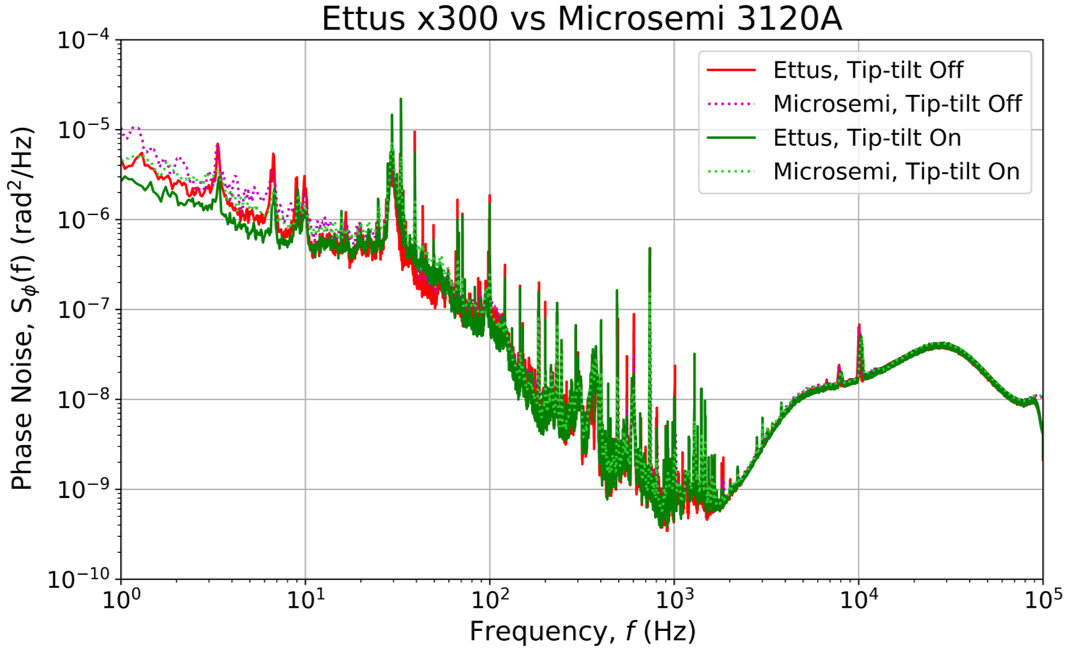}
    \caption{Concurrent measurements of 265~m free-space link phase noise with tip-tilt active optics off (red) and on (green) with an Ettus X300 (solid lines) and Microsemi 3120A (dotted lines). The Ettus and Microsemi are in very close agreement across the Microsemi's frequency range.}
    \label{fig:ettusVsMicro}
\end{figure}

\newpage
\section{Noise contributions in parallel compensated links}\label{sec:noiseFloor}
Here, we estimate the noise floor of the phase noise measurements presented in Section~\ref{sec:results}. The aim is to estimate the phase stabilization noise floor caused by atmospheric noise and laser frequency noise when compensating the two links.

\subsection{Differential noise PSD}
For a noise process $x(t)$ that is stationary and characterised by a PSD $S_{x}(f)$, the PSD of the linear combination $z(t) \equiv \sum {a_i x(t-T_i)}$ (where $a_i$ are arbitrary real constants) can be derived from first principles (see e.g. Duchayne \citeAppendix{Duchayne} p.227) and is given by
\setcounter{equation}{0}
%
%
%

\begin{equation} \label{equ:S_gen2}
S_z(f) = \left(\sum\limits_i a_i^2 + \sum\limits_i \sum\limits_{j\neq i} a_ia_j {\rm cos}(2\pi f (T_i-T_j))\right)S_x(f) .
\end{equation}

\subsection{A single compensated link}
Assume a compensated link (either of the two in Fig.~\ref{fig:experimentalSystem}) where a laser signal is frequency shifted by a transmission AOM before being injected into the link. 

The frequency of the laser signal is $\nu_{L}(t) = \nu_L + \Delta\nu_{L}(t)$ where $\Delta\nu_{L}(t)$ is the laser frequency noise. The transmission AOM shifts the optical frequency by a nominal frequency ($\nu_{tr}$) which is varied by $\Delta\nu_{tr}(t)$ in order to suppress link noise. $\Delta\nu_{tr}(t)$ is ideally equal but opposite to the frequency shift that the signal will experience during propagation through the link ($\delta\nu(t)$). At the remote site, the optical signal is passed through a static anti-reflection AOM ($\nu_{ar}$) before being passed out of the remote site. We will define $t$ as the instant the signal is passed out of the remote site. The laser signal coming out at the remote site has frequency $\nu_{out}(t)$.

\begin{equation}\label{equ:uncompensated}
\nu_{out}(t)=\nu_L +\Delta\nu_{L}(t-T)+\nu_{tr}+\Delta\nu_{tr}(t-T)+\nu_{ar}+\delta\nu(t),
\end{equation}

\noindent where $T$ is the propagation time through the link.

The frequency shift $\delta\nu(t)$ is compensated by measuring, at the local end, the beatnote between $\nu_{L}(t)$ and a return signal ($\nu_{ret}(t)$) that went through the AOM twice (for simplicity we will ignore the remote site anti-reflection AOM, $\nu_{ar}$, and the nominal transmission AOM frequency, $\nu_{tr}$, applied by the transmission AOM):

\begin{eqnarray}\label{equ:beatnote}
\nu_{L}(t)-\nu_{ret}(t) &=&\nu_{L}(t) - \left[\nu_{L}(t-2T) + \Delta\nu_{tr}(t-2T)+\delta\nu(t-T)+\delta\nu(t)+\Delta\nu_{tr}(t)\right]\nonumber \\
&=& \Delta\nu_{L}(t) - \left[\Delta\nu_{L}(t-2T) + \Delta\nu_{tr}(t-2T)+\delta\nu(t-T)+\delta\nu(t)+\Delta\nu_{tr}(t)\right]\nonumber \\
\end{eqnarray}

The phase-locked loop (PLL) \footnote{We assume that the delay in the PLL is negligible with respect to $T$.} ensures that the beatnote signal (\ref{equ:beatnote}) is zero, which means 

\begin{eqnarray}
\Delta\nu_{tr}(t) +\Delta\nu_{tr}(t-2T)= (\Delta\nu_{L}(t) - \Delta\nu_{L}(t-2T)) -(\delta\nu_{L}(t-T)+\delta\nu_{L}(t))\, . 
\end{eqnarray}

We linearize the expression by a Taylor expansion of the left side leading to a differential equation: 

\begin{eqnarray}\label{equ:nuAOM_DE}
\Delta\nu_{tr}(t) -T\Delta\dot{\nu}_{tr}(t)\simeq \frac{1}{2}(\Delta\nu_{L}(t) - \Delta\nu_{L}(t-2T)) -(\delta\nu(t-T)+\delta\nu(t))\, ,  
\end{eqnarray}

\noindent where we neglect higher order terms in the Taylor expansion. The general solution of the differential equation is

\begin{eqnarray}
\Delta\nu_{tr}(t) \simeq Ce^{t/T} + \frac{1}{2}(\Delta\nu_{L}(t) - \Delta\nu_{L}(t-2T)) -(\delta\nu(t-T)+\delta\nu(t))\, , 
\end{eqnarray}

\noindent where $C$ is an integration constant. As our loop is closed (the AOM frequency does not diverge) we have $C=0$. Substituting that into (\ref{equ:uncompensated}) we finally get

\begin{eqnarray}\label{equ:final_out}
\nu_{out}(t)&=&\nu_{L} +\Delta\nu_{L}(t-T)+\delta\nu(t) + \frac{1}{2}(\Delta\nu_{L}(t-T) - \Delta\nu_{L}(t-3T)) -(\delta\nu(t-2T)+\delta\nu(t-T))\, .
\end{eqnarray}

As one would expect, the expression reduces to $\nu_{out}(t)=\nu_L +\Delta\nu_{L}(t)$ when laser and link noise vary slowly with respect to $T$, i.e. the output signal is a copy of the input signal in spite of the presence of link noise.

Applying (\ref{equ:S_gen2}) to the laser noise contributions gives
\begin{equation} \label{equ:S_single_laser}
S_{out}(f) = S_{\Delta\nu}(f) + \frac{3}{2}\Big(1-{\rm cos}(4\pi f T)\Big)S_{\Delta\nu}(f).
\end{equation}
The second term quickly vanishes for $f \ll 1/T$ and we just have the input laser noise at the output. At approximately $f\geq 1/T$ the noise is increased by the compensation by a frequency dependent term that oscillates between 0 and $3S_{\Delta\nu}(f)$.

For the link noise, we obtain also from (\ref{equ:S_gen2})
\begin{equation} \label{equ:S_single_link}
S_{out}(f) = \Big(\frac{3}{2}-\frac{1}{2}{\rm cos}(2\pi f T)-{\rm cos}(4\pi f T)\Big)S_{\delta\nu}(f)\,,
\end{equation}
which, as expected, cancels completely when $f \ll 1/T$.

The total noise PSD of the output signal is the sum of (\ref{equ:S_single_laser}) and (\ref{equ:S_single_link}).

\subsection{Two parallel compensated links}\label{sec:twoLinks}
We now consider two parallel compensated links A and B with delays $T_A$ and $T_B$ that are fed by the same input laser (see Figure~\ref{fig:experimentalSystem}). We will be interested in the difference $D\nu_{out}(t) \equiv \nu_{outA}(t) - \nu_{outB}(t)$, which is measured in the experiment. Each link is affected by the same laser noise $\Delta\nu_{L}(t)$, but by different link noises $\delta\nu_A(t)$ and $\delta\nu_B(t)$.  

For the laser noise applying (\ref{equ:final_out}) we directly have
\begin{eqnarray}\label{equ:TwoLinks}
D\nu_{out}(t)&=&\frac{3}{2}\Delta\nu_{L}(t-T_A)-\frac{1}{2}\Delta\nu_{L}(t-3T_A) \nonumber \\
&-&\frac{3}{2}\Delta\nu_{L}(t-T_B)+\frac{1}{2}\Delta\nu_{L}(t-3T_B).
\end{eqnarray}
When the two links have the same delay ($T_A=T_B$) the laser noise cancels exactly. When they are different one can apply (\ref{equ:S_gen2}) to calculate the overall effect.

We assume that the link noise is uncorrelated between the two links, and thus the link PSDs given by (\ref{equ:S_single_link}) simply add.

The final result is then
\begin{eqnarray}\label{TwoLinksS2}
S_{D\nu}(f) &=& S_{\Delta\nu_{L}}(f)\Bigg(5 - \frac{9}{2}{\rm cos}(2\pi f \Delta T) - \frac{1}{2}{\rm cos}(6\pi f \Delta T)  \\
&-& 6 \Big({\rm cos}(4\pi f T_A) + {\rm cos}(4\pi f T_B) + {\rm cos}(2\pi f (T_A+T_B))\Big){\rm sin}^2\left(\pi f \Delta T \right)\Bigg)\nonumber \\
&+& S_{\delta\nu A}(f)\Big(\frac{3}{2}-\frac{1}{2}{\rm cos}(2\pi f T_A)-{\rm cos}(4\pi f T_A)\Big) \nonumber\\
&+& S_{\delta\nu B}(f)\Big(\frac{3}{2}-\frac{1}{2}{\rm cos}(2\pi f T_B)-{\rm cos}(4\pi f T_B)\Big)\, \nonumber.
\end{eqnarray}
\noindent where $\Delta T \equiv T_A-T_B$. As expected from (\ref{equ:TwoLinks}), the laser noise contribution vanishes for $T_A=T_B$. In the limit when $f\ll 1/T$ the link noise contributions vanish, and (\ref{TwoLinksS2}) can be approximated as
\begin{equation}
S_{D\nu}(f) \simeq S_{\Delta\nu_{L}}(f)\,36\,\pi^4(T_A^2-T_B^2)^2 f^4 \,.
\end{equation}

\subsection{Noise floor estimation}
To estimate the noise floor in our compensated link we need to evaluate (\ref{TwoLinksS2}) using estimates of the free space link noise $S_{\delta\nu A}(f)$, the fiber link noise $S_{\delta\nu B}(f)$, and the laser noise $S_{\Delta\nu_{L}}(f)$. We assume that the fiber link noise is negligible with respect to the atmospheric noise and use the ``unstabilized, tip-tilt off" measurement (see fig. \ref{fig:stabilityResults}) as our estimate of the latter. For the laser noise we use the ``typical" noise curve given by the manufacturer \citeAppendix{NKT}. The two delays were estimated from delay measurements in the fiber and from local maps as $cT_A \approx 302$~m (265~m free space + 25~m fibers with refractive index $n=1.65$ between the telescopes and the beam splitters), and $cT_B = n\times 715$~m. The result is shown on figure \ref{fig:noiseFloor}.

\setcounter{figure}{0}
\begin{figure}[ht]
    \centering
    \includegraphics[width=0.6\linewidth]{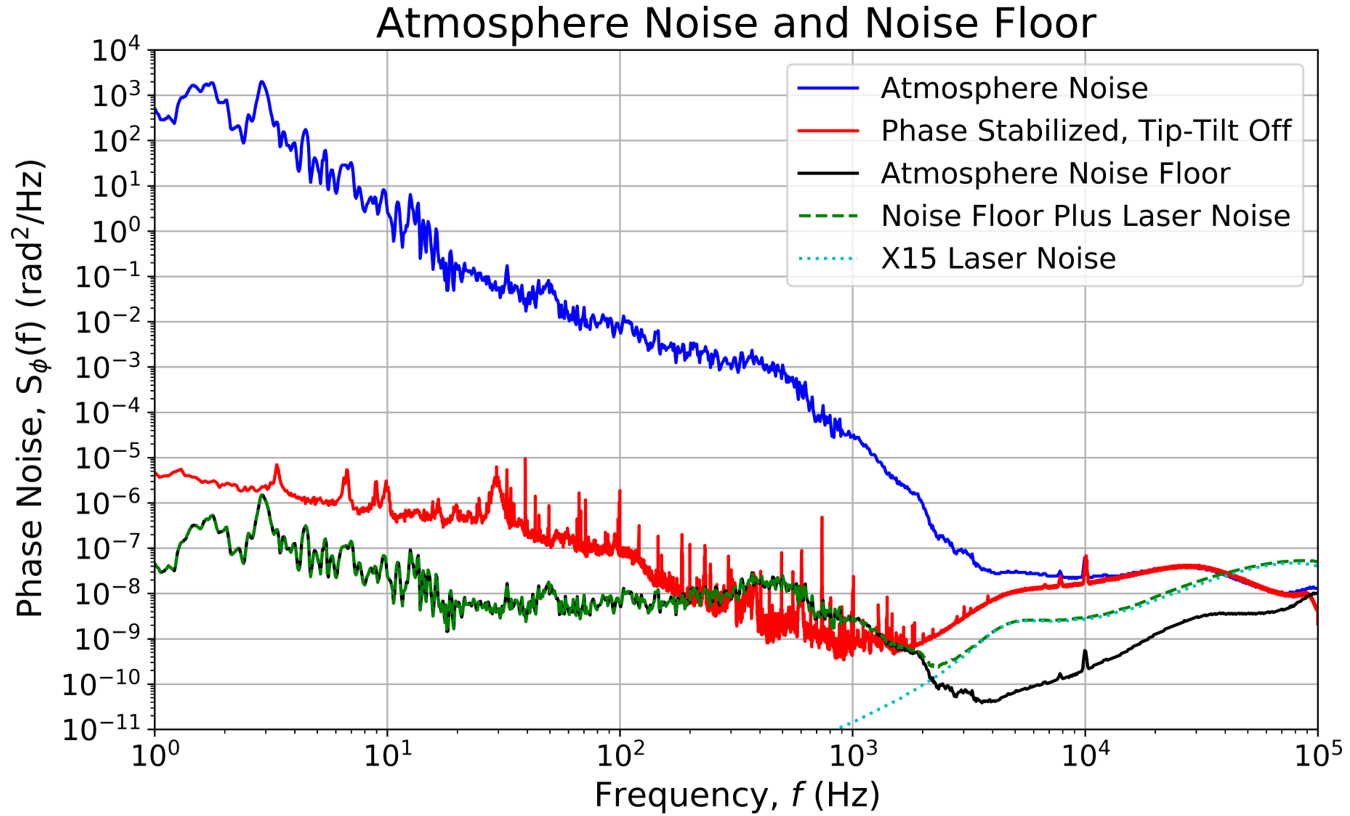}
    \caption{Estimate of the noise floor for the parallel compensated links. Blue, free-running noise from Figure~\ref{fig:stabilityResults} dominated by atmospheric phase noise; red, phase stabilized, tip-tilt off result from Figure~\ref{fig:stabilityResults}; black, estimate of noise floor due to atmosphere; dashed green, atmospheric noise floor plus estimate of laser noise floor over the parallel links; dotted cyan, estimate of laser noise floor.}
    \label{fig:noiseFloor}
\end{figure}

We note that above about 200~Hz our measured results are well explained by the combined effects of atmospheric turbulence and laser noise, at least in terms of the orders of magnitude. The residual noise of the compensated links is dominated by atmospheric noise between 200~Hz and 2~kHz, and laser noise takes over at higher frequencies. The slight discrepancy in the atmospheric effect is probably due to the fact that the uncompensated measurement (blue line in fig. \ref{fig:noiseFloor}) was taken at a different time than the compensated one. The discrepancy at higher frequency is most likely due to the fact that the NKT X15 laser we used had more phase noise than the ``typical" performance of that type of laser specified by the manufacturer in \citeAppendix{NKT}. Nonetheless the similar shape and the fact that all measured curves on Fig.~\ref{fig:stabilityResults} converge above 10~kHz comforts us in the hypotheses that laser noise is the common origin. The exception is the ``system noise floor" shown in black on Fig.~\ref{fig:stabilityResults} for which two parallel fibres of equal length were used i.e. $T_A=T_B$ in (\ref{TwoLinksS2}) and laser noise cancels, as can be observed.

\bibliographystyleAppendix{unsrt}

\end{appendices}
\end{document}